\documentclass[aoas,MSNbibl,nameyear,noautosecdot,dvips]{arximspdf}
\usepackage{dcolumn}
\usepackage{graphicx}

\doi{10.1214/14-AOAS770} % kopijuoti is laisko
\volume{8}
\issue{4}
\pubyear{2014}
\firstpage{2096}
\lastpage{2121}
\docsubty{FLA}

\makeatletter
\newcolumntype{d}[1]{D{.}{.}{#1}}
\newcommand{\rrvert}{\vert}
\newcommand{\llvert}{\vert}
\makeatother

\begin{document}
\begin{frontmatter}

\title{Isolation in the construction of natural experiments}
\runtitle{Isolation in the construction of natural experiments}
\begin{aug}
\author[A]{\fnms{Jos\'{e} R.}~\snm{Zubizarreta}\corref{}\thanksref{M1}\ead[label=e1]{zubizarreta@columbia.edu}},
\author[B]{\fnms{Dylan S.}~\snm{Small}\thanksref{T2,M2}\ead[label=e2]{dsmall@wharton.upenn.edu}}
\and
\author[B]{\fnms{Paul R.}~\snm{Rosenbaum}\thanksref{T2,M2}\ead[label=e3]{rosenbaum@wharton.upenn.edu}}
\thankstext{T2}{Supported by NSF Grant SES-1260782.}
\runauthor{J.~R. Zubizarreta, D.~S. Small and P.~R. Rosenbaum}
\affiliation{Columbia University\thanksmark{M1} and University of
Pennsylvania\thanksmark{M2}\vspace*{6pt}}
\address[A]{J. R. Zubizarreta\\
Division of Decision, Risk and Operations\\
and\\
Department of Statistics\\
Columbia University\\
New York, New York 10027\\
USA\\
\printead{e1}}
%adresu isvedimo komanda gale!
\address[B]{D.~S. Small\\
P.~R. Rosenbaum\\
Department of Statistics\\
The Wharton School\\
University of Pennsylvania\\
Philadelphia, Pennsylvania 19104-6340\\
USA\\
\printead{e2}\\
\phantom{E-mail:\ }\printead*{e3}}
\end{aug}

% HISTORY:
\received{\smonth{10} \syear{2013}}
\revised{\smonth{7} \syear{2014}}

% ABSTRACT
%
\begin{abstract}
A natural experiment is a type of observational study in which
treatment assignment, though not randomized by the investigator, is
plausibly close to random. A process that assigns treatments in a highly
nonrandom, inequitable manner may, in rare and brief moments, assign aspects
of treatments at random or nearly so. Isolating those moments and aspects
may extract a natural experiment from a setting in which treatment
assignment is otherwise quite biased, far from random. Isolation is a tool
that focuses on those rare, brief instances, extracting a small natural
experiment from otherwise useless data. We discuss the theory behind
isolation and illustrate its use in a
reanalysis of a well-known study of the effects of fertility on workforce
participation. Whether a woman becomes pregnant at a certain moment in her
life and whether she brings that pregnancy to term may reflect her
aspirations for family, education and career, the degree of control she
exerts over her fertility, and the quality of her relationship with the
father; moreover, these aspirations and relationships are unlikely to be
recorded with precision in surveys and censuses, and they may confound
studies of workforce participation. However, given that a women is
pregnant and will bring the pregnancy to term, whether she will have twins
or a single child is, to a large extent, simply luck. Given that a woman
is pregnant at a certain moment, the differential comparison of two
types of
pregnancies on workforce participation, twins or a single child, may be
close to randomized, not biased
by unmeasured aspirations. In this comparison, we find in our case
study that
mothers of twins had more children but only slightly reduced workforce
participation, approximately 5\% less time at work for an additional
child.\vspace*{12pt}
\end{abstract}

% KEYWORDS
\end{frontmatter}

%s1 #&#
\section{\texorpdfstring{Constructing natural experiments.}{Constructing natural experiments}}
\label{secIntro}

%s1.1 #&#
\subsection{\texorpdfstring{Natural experiments.}{Natural experiments}}

Natural experiments are a type of observational study, that is, a study of
the effects caused by treatments when random assignment is infeasible or
unethical. What distinguishes a natural experiment from other
observational studies is the emphasis placed upon finding unusual
circumstances in which treatment assignment, though not randomized,
seems to
resemble randomized assignment in that it is haphazard, not the result of
deliberation or considered judgement, not confounded by the typical
attributes that determine treatment assignment in a particular empirical
field.\vadjust{\goodbreak} The literature on natural experiments spans the health and social
sciences; see, for instance, \citet{ArpAas13}, Imai et al. (\citeyear{Imaetal11}), \citet{Mey95}, \citet{Rut07}, \citet
{SekTit12}, \citet{Sus81} and \citet{Van04}.

Traditionally, natural experiments were found, not built. In one sense,
this seemed inevitable: one needs to find haphazard treatment
assignment in
a world that typically assigns treatments in a biased fashion, often
assigning treatments with a view to achieving an objective. There is,
however, substantial scope for constructing natural experiments. When
treatment assignment is biased, there may be aspects of treatment
assignment, present only briefly, that are haphazard, close to random. The
key to constructing natural experiments is to isolate these transient,
haphazard aspects from typical treatment assignments that are biased. If
brief haphazard aspects of treatment assignment can be isolated from the
rest, in the isolated portion it is these haphazard elements that are
decisive. This is analogous to a laboratory in which a treatment is
studied in isolation from disruptions that would obscure the treatment's
effects. Laboratories are built, not found.

%s1.2 #&#
\subsection{\texorpdfstring{A natural experiment studying effects of fertility on workforce
participation.}{A natural experiment studying effects of fertility on workforce
participation}}
\label{ssAEexample}

Does having a child reduce a mother's participation in the workforce?
If it does, what is
the magnitude of the reduction? The question is relevant to individuals
planning families
and careers and to legislators and managers who determine policies
related to fertility, such as
family leaves. A major barrier to answering this question is that, for
many if not most
women, decisions about fertility, education and career are highly
interconnected, and each
decision has consequences for the others. Here we follow Angrist and Evans (\citeyear{AngEva96})
and seek to determine if
there is some source of variation in fertility that does not reflect
career plans and is just luck. Although a woman has the ability to
influence the timing of her pregnancies, given that she is pregnant at
a particular age,
she has much less influence about whether she will have a boy or a
girl, whether she will
have a single child or twins---to a large extent, that is just luck.
More precisely, that
a woman is pregnant at a certain moment in her life may be indicative
of her unrecorded
plans and aspirations for education, family and career, but
conditionally given that she is
pregnant at that moment, the birth outcome, a~boy or a girl or twins,
is unlikely to indicate
much about her plans and aspirations.

We focus here on the haphazard contrast most likely to shift the total
number of children, namely, a comparison of similar women, one with a
twin at her $k$th birth, the other with children of mixed sex at her
$k$th birth since, as Angrist and Evans (\citeyear{AngEva96}) noted, many
women or families in the US prefer to have children of both sexes,
rather than just boys or
just girls, that is, a third child is seen in data to be more common if
the first two children
have the same sex. While we could compare women having twins with women
having a single child whose sex is the same as her first child, we
focus on comparing women having twins with women having a single child
whose sex is different from her first since the first woman may end up
with one more child than she intended, whereas the other woman will, at
least, not have additional children simply to have one of each sex.\vadjust{\goodbreak}

What question does such a natural experiment answer? Conditionally
given that a
woman with a certain prior history of fertility is currently pregnant,
having a girl or a boy
or twins does not pick out a particular type of woman. So the study is
accepting whatever
process led a particular woman to be pregnant at a certain moment in
her life, and it is
asking: What would happen if she unexpectedly had two children at that
pregnancy rather than one? How would that event alter her subsequent
workforce participation? We use the idea from Angrist and Evans (\citeyear{AngEva96})
to illustrate and discuss tools to extract
natural experiments from larger biased data sets, in particular, risk
set matching [Li, Propert and Rosenbaum (\citeyear{LiProRos01})], differential effects
[Rosenbaum (\citeyear{Ros06,Ros13N1})] and strengthening an
instrumental
variable [\citet{Baietal10}, Zubizarreta et al. (\citeyear{Zubetal13})].

%s1.3 #&#
\subsection{\texorpdfstring{Informal review of two key concepts: Differential effects;
risk-set\break matching.}{Informal review of two key concepts: Differential effects;
risk-set matching}}
\label{ssInformalReview}

Because differential effects and risk set matching may be unfamiliar,
we now
review the motivation for these techniques. Consider, first, differential
effects and generic biases acting at a single point in time [Rosenbaum
(\citeyear{Ros06,Ros13N1})]. Treatment assignment may be biased by certain unmeasured
covariates that promote several treatments in a similar way. When this is
true, receiving a treatment $s$ may be very biased by these covariates,
while receiving one treatment $s$ in lieu of another $s^{\prime}$ may be
unbiased or less biased or biased in a different way. Here, attention
shifts from whether or not a person received treatment $s$ (i.e., the main
effect of~$s$) to whether a person received treatment $s$ rather than
treatment $s^{\prime}$ conditionally given that the person received either
treatment $s$ or treatment $s^{\prime}$ (i.e., the differential effect
of $%
s $ in lieu of $s^{\prime}$). Consider an example discussed in detail by
\citet{Antetal00}. There is a theory that nonsteroidal
anti-inflammatory drugs (NSAIDs), such as ibuprofen (e.g., brand
Advil), may
reduce the risk of Alzheimer disease. There is an obvious bias in
comparing people who regularly take ibuprofen and people who do not. In
all likelihood, a person who regularly takes ibuprofen is experiencing
chronic pain, perhaps arthritis or back pain, is aware of that pain,
and is
capable of acting deliberately on the basis of that awareness. It has been
suggested that people in the early undiagnosed stages of Alzheimer disease
are less aware of pain and less able to act on what awareness they
have, so
that fact alone might produce a spurious association between use of
ibuprofen and lower risk of later diagnosed Alzheimer disease. There are,
however, pain relievers that are not NSAIDs, for example, acetaminophen
(e.g., brand Tylenol). While limited awareness of pain or limited ability
to act on awareness might reduce use of pain relievers of all kinds, it
seems far less plausible that it shifts people away from ibuprofen and
toward acetaminophen. That is, the differential effect of
acetaminophen-versus-ibuprofen---of one treatment in lieu of the
other---may not be biased by unmeasured covariates that would bias straightforward
estimates of the main effect of either drug. Differential effects are not
main effects, but when differential effects are interesting, they may be
immune to certain biases that distort main effects. See also \citet
{Gibetal10} for differential effects in the study of medications.

Consider, second, risk-set matching, a device for respecting the temporal
structure of treatment assignment in observational studies [Li, Propert and Rosenbaum (\citeyear{LiProRos01})].
For each subject in a randomized experiment, there is a specific
moment at
which this subject is assigned to treatment or to control. In some
observational studies, there is no corresponding moment. Some people
receive treatment at a specific time, others receive it later or never
receive it, but anyone who does not receive treatment today might
receive it
tomorrow. Risk-set matching pairs two individuals at a specific time, two
individuals who looked similar in observed covariates prior to that specific
time, a time at which one individual was just treated and the other was
not-yet-treated. The not-yet-treated individual may be treated tomorrow,
next year or never. We compare two individuals who looked similar prior
to the moment that one of them was treated, avoiding matching or adjustment
for events subsequent to that moment [cf. \citet{Ros84}]. That is, in
the language of Cox's proportional hazards model, risk-set matching pairs
two individuals who were both at risk of receiving the treatment a moment
before one of them actually received it, two individuals who looked similar
in time-dependent covariates prior to that moment. Taken alone, without
differential comparisons, risk-set matching is a method for controlling
measured time-dependent covariates respecting the temporal structure of
treatment assignment; see \citet{vanRob03} for other methods
for this task.

%s1.4 #&#
\subsection{\texorpdfstring{Outline of the paper.}{Outline of the paper}}

Section~\ref{secRiskSetGeneric} discusses new relevant theory, specifically
theory linking risk-set matching for time-dependent measured covariates with
differential comparisons unaffected by certain unmeasured time dependent
covariates. Fertility is commonly modeled in terms of ``event history'' or point process models determining the
timing of events together with ``marks'' or
random variables describing these randomly timed events. The mark may
record the occurrence of twins. Temporal order is key and must be
respected. Sections~\ref{secAEconstruction}
and \ref{secAnalysis} complete the case study of twin births with the
construction of the matched sample using combinatorial optimization for
risk-set matching discussed in Section~\ref{secAEconstruction} and a
detailed analysis presented in Section~\ref{secAnalysis}. Section~\ref{sec5}
includes a discussion of related work.

%s2 #&#
\section{\texorpdfstring{Risk-set matching to control generic unmeasured biases.}{Risk-set matching to control generic unmeasured biases}}
\label{secRiskSetGeneric}

%s2.1 #&#
\subsection{\texorpdfstring{Notation for treatments over time.}{Notation for treatments over time}}
\label{ssNotation}

The population before matching contains statistically independent
individuals. At time $t$, individual $\ell$ has a history of events prior
to $t$, the observed history being recorded in $\mathbf{x}_{\ell t}$
and the
unobserved history being recorded in $u_{\ell t}$. To avoid a formal
notation that we would rarely use, we write histories as variables,
$\mathbf{%
x}_{\ell t}$ or $u_{\ell t}$, but we intend to convey a little more than
this. Both the quantity and types of information in $\mathbf{x}_{\ell t}$
or in $u_{\ell t}$ or in $ ( \mathbf{x}_{\ell t},u_{\ell t} ) $
increase as time passes, that is, as $t$ increases [or, formally, the sigma
algebra generated by $ ( \mathbf{x}_{\ell t},u_{\ell t} ) $ is
contained within the sigma algebra generated by $ ( \mathbf{x}_{\ell
t^{\prime}},u_{\ell t^{\prime}} ) $ for $t<t^{\prime}$].

In our case study, $\mathbf{x}_{\ell t}$ records such things as the
ages at
which mother $\ell$ gave birth to the children she had prior to time $t$,
her years of education attained at the times of those births before
time $t$%
, and unchanging characteristics such as her place of birth, race or
ethnicity. In parallel, $u_{\ell t}$ might be an unmeasured quantity
reflecting the entire history of a woman's inclination to work full
time in
the year subsequent to time $t$. Obviously, a birth at time $t$ might,
often would, alter $\mathbf{x}_{\ell t^{\prime}}$ or $u_{\ell
t^{\prime}}$
for $t^{\prime}>t$.

There is also a treatment process $Z_{\ell t}$ that is in one of $K+1$
states, $s_{0}$, $s_{1}, \ldots, s_{K}$. That is, at any time $t$,
individual $\ell$ is in exactly one of these states, $Z_{\ell
t}=s_{k}$ for
some $k\in \{ 0,1,\ldots,K \} $. Also, write $\overline{Z}%
_{\ell t}$ for the history of the $Z_{\ell t}$ process strictly prior to
time $t$, so $\overline{Z}_{\ell t}$ records $Z_{\ell t^{\prime}}$ for
$%
t^{\prime}<t$ but it does not record $Z_{\ell t}$. In our case study,
state $s_{0}$ is the interval state of not currently giving birth to a
child, state $s_{1}$ is the point state of giving birth to a single female
child, state $s_{2}$ is the point state of giving birth to a single male
child, state $s_{3}$ is the point state of giving birth to a pair of female
twins and so on. Most women are in state $Z_{\ell t}=s_{0}$ at most times
$t$. The history $\overline{Z}_{\ell t}$ records mother $\ell$'s
births up
to time $t$, their timing, the sex of the child, twins, etc.

Consider a specific individual $\ell$ at a specific time $t$. At this
moment, the individual has a treatment history $\overline{Z}_{\ell t}$ prior
to $t$ and is about to receive a current treatment~$Z_{\ell t}$. Given the
past, $\overline{Z}_{\ell t}$, we are interested in the effect of the
current treatment $Z_{\ell t}$ on some future (i.e., after $t$) outcome
$%
R_{\ell}$. Write $\mathcal{F}_{\ell t}= ( \overline{Z}_{\ell t}, %
\mathbf{x}_{\ell t}, u_{\ell t} ) $ for the past at time $t$. In
parallel with Neyman (\citeyear{Ne23}) and \citet{Rub74}, this individual
$\ell$ at
this time $t$ has $K+1$ possible values for $R_{\ell}$ depending upon the
treatment $Z_{\ell t}$ assigned at time $t$, that is, $R_{\ell
}=r_{k\ell}$
if $Z_{\ell t}=s_{k}$, where only one $R_{\ell}$ is observed from an
individual, and the effect of giving treatment $k$ rather than
$k^{\prime}$
at time $t$, namely, $r_{k\ell}-r_{k^{\prime}\ell}$ is not observed for
any person at any time. This structure is for individual $\ell$ at a
specific time $t$ with treatment history $\overline{Z}_{\ell t}$; typically,
everything about this structure would change if the history $\overline
{Z}%
_{\ell t}$ to time $t$ had been different. The question is what effect
treatment at time $t$ has on an individual with a specific treatment and
covariate history prior to $t$. It is entirely possible---indeed, in
typical applications, it is likely---that the treatments $Z_{\ell
t^{\prime}}$ at times $t^{\prime}<t$ alter the value of observed or
unobserved subsequent history $ ( \mathbf{x}_{\ell t},u_{\ell t} ) $%
, but the history at $t$, namely, $ ( \mathbf{x}_{\ell t},u_{\ell
t} ) $, records the situation just prior to $t$ and hence is unaffected
by the treatment assignment $Z_{\ell t}$ at $t$. Quite often, the
outcome $%
R_{\ell}$ is a future value of a quantity that is analogous to a past
quantity recorded in the history $ ( \mathbf{x}_{\ell t},u_{\ell
t} ) $. In our case study, $R_{\ell}$ might measure an aspect of
future workforce participation beyond time $t$ where $ ( \mathbf{x}%
_{\ell t},u_{\ell t} ) $ records workforce participation prior to time $
t$, or $R_{\ell}$ might measure educational attainment at some time
after $t
$ where $ ( \mathbf{x}_{\ell t},u_{\ell t} ) $ records educational
attainment prior to time $t$.

In our case study, aspects of the record of a woman's fertility,
$Z_{\ell t}$%
, are likely to be strongly predicted by aspects of her observed and
unobserved histories $ ( \mathbf{x}_{\ell t},u_{\ell t} ) $. A
woman $\ell$ aged $t^{\prime}=18$ years whose private aspiration
$u_{\ell
t}$ is to earn a Ph.D. in molecular biology and an MBA and to start her own
biotechnology company is likely to take active steps to ensure $Z_{\ell
t}=s_{0}$ for $t\in ( 18,22 ] $ or longer, that is, she is likely
to postpone having children for at least several years. In contrast,
another woman $\ell^{\prime}$ whose private aspiration $u_{\ell
^{\prime
}t}$ at age $t^{\prime}=18$ is to stay at home as the mother of many
children may take active steps to ensure $Z_{\ell t}\neq s_{0}$ for
several $%
t\in ( 18,22 ] $, that is, she may actively pursue her goal of a
large family. A comparison of the workforce participation of woman
$\ell$
and woman $\ell^{\prime}$ will be severely biased as an estimate of the
effects of having a child before age 22 on workforce participation,
because $%
\ell$ tried to shape her fertility to fit her work plans and $\ell
^{\prime
}$ tried to shape her fertility to fit her family plans---even if, by some
accident, they had the same pattern of fertility over $t\in ( 18,22%
] $, we would not be surprised to learn that $\ell$ subsequently
worked more for pay than did $\ell^{\prime}$. What is an investigator to
do when unmeasured aspirations, intentions and goals are strongly associated
with treatment assignment?

%s2.2 #&#
\subsection{What is risk-set matching?}
\label{ssWhatIsRiskSetMatching}

Risk-set matching compares people, say, $\ell$ and $\ell^{\prime}$, who
received different treatments at time $t$, $Z_{\ell t}\neq Z_{\ell
^{\prime
}t}$, but who looked similar in their observed histories prior to $t$, $
\mathbf{x}_{\ell t}=\mathbf{x}_{\ell^{\prime}t}$ and $\overline
{Z}_{\ell
t}=\overline{Z}_{\ell^{\prime}t}$; see Li, Propert and Rosenbaum (\citeyear{LiProRos01}), \citet
{Lu05} and
Rosenbaum [(\citeyear{Ros10}), Section~12]. Importantly, $\ell$ and
$\ell^{\prime}$ are
similar prior to $t$ in terms of observable quantities that may be
controlled by matching, but they may not be similar in terms of unmeasured
histories, $u_{\ell t}\neq u_{\ell^{\prime}t}$, and of course they may
differ in the future, after time $t$, not least because they received
different treatments at time $t$. Risk-set matching does not solve the
problem of unmeasured histories. Risk-set matching does respect the
temporal structure of the data, avoiding adjustment for variables affected
by the treatment [\citet{Ros84}]. Risk-set matching also
``simplifies the conditions of
observation,'' to use Mervyn Susser's [(\citeyear
{Sus73}), Section~7]
well-chosen phrase, ensuring that comparisons are of people with histories
that look comparable, even though those histories may be of different
lengths, and hence may contain qualitatively different information.
Although individuals have histories of different lengths containing
qualitatively different information, matched individuals have histories of
the same length. For instance, a woman giving birth to her 3rd child
has in her history ages of birth of her first three children, where a mother
giving birth to her second child does not have in her history her age
at the
birth of her third child, if indeed she had a third child.

In implementing risk-set matching in Section~\ref{secAEconstruction},
we match
women of the same age, with the same history of fertility---the same
numbers of prior children born at the same ages in the same patterns. We
also control for temporally fixed quantities associated with fertility, such
as ethnicity. A delicate issue that risk-set matching would
straightforwardly address with adequate data is
``education.'' On the one hand, education is strongly
related to wage income and is related to employment, so it may strongly
predict certain workforce outcomes $R_{\ell}$. On the other hand,
education may itself be affected by fertility: a mother who has her first
child at age 16 may as a consequence have difficulty completing high school.
In principle, the issue is straightforward with risk-set matching: in
studying the effects of fertility $Z_{\ell t}$ at time $t$, one
compares two
people who had the same education prior to $t$, without equating their
educations subsequent to time $t$. Again, this avoids adjustment for
variables affected by the treatment [\citet{Ros84}]. If the adjustment
for education at time $t$ controlled for subsequent education at time $%
t^{\prime}>t$, it might---probably would---remove a substantial
part of
the actual effect on workforce participation of having a child at age
16. Not finishing high school is a good way to have trouble in the
labor market,
and having a child at age 16 is a good way to have trouble finishing high
school; everyone remembers this until they start running regressions, but
then, too often, part of an actual effect is removed by adjusting for a
posttreatment variable that was also affected by the treatment.

Risk-set matching was discussed by Li, Propert and Rosenbaum (\citeyear{LiProRos01}) and \citet
{Lu05}. It has
been applied in criminology [Nieuwbeerta, Nagin and
Blokland (\citeyear{NieNagBlo09}), \citet{Apeetal10},
Murray, Loeber and Pardini (\citeyear{MurLoePar12})], sociology [Wildeman, Schnittker and
Turney (\citeyear{WilSchTur12})] and medicine
[\citet{Kenetal10}]. See \citet{Maretal08},
Rosenbaum [(\citeyear{Ros10}), Section~12], \citet{Stu10} and
\citet{Luetal11} for related discussion.

%s2.3 #&#
\subsection{\texorpdfstring{Removing generic unmeasured biases by differential comparisons
in risk sets.}{Removing generic unmeasured biases by differential comparisons
in risk sets}}
\label{ssDifRiskMatch}

The model for biased treatment assignment in risk-set matching is intended
to express the thought that matching for the observed past, $ (
\overline{Z}_{\ell t}, \mathbf{x}_{\ell t} ) $, has controlled for the
observed past but typically did not control for the unobserved past~$u_{\ell
t}$. The model is a slight generalization to multiple states of the model
for two states in Li, Propert and Rosenbaum [(\citeyear{LiProRos01}), Section~4], and that model is itself closely
patterned after Cox's (\citeyear{Cox72}) proportional hazards model for
outcomes rather
than treatments. People are in state $s_{0}$ almost all the time, and are
in states $s_{1},\ldots,s_{K}$ only at points in time. Let $\lambda
_{k} ( \mathcal{F}_{\ell t} ) =\lambda_{k} ( \overline{Z}%
_{\ell t}, \mathbf{x}_{\ell t}, u_{\ell t} ) $ be the hazard, assumed
to exist, of entering state $k\geq1$ at time $t$ given past $\mathcal
{F}%
_{\ell t}$. The hazard is assumed to be of the form $\lambda_{k} (
\overline{Z}_{\ell t}, \mathbf{x}_{\ell t}, u_{\ell t} ) =\exp
\{ \kappa_{k} ( \overline{Z}_{\ell t}, \mathbf{x}_{\ell t} )
+\phi_{k} u_{\ell t} \} $ where $\kappa_{k} ( \cdot, \mathbf{%
\cdot} ) $ is unknown. Because $\mathbf{x}_{\ell t}$ may include as
one of its coordinates the time $t$, this model permits the hazards to vary
with time $t$. For state $s_{0}$, it is notationally convenient to
define $%
\lambda_{0} ( \cdot, \mathbf{\cdot}, \cdot ) =1$ and $\phi
_{0}=0$.

In Section~\ref{ssNotation}, $u_{\ell t}$ was described as a possibly
multivariate history of a possibly continuous process in time, whereas in
the hazard model,\break $\exp \{ \kappa_{k} ( \overline{Z}_{\ell t}, %
\mathbf{x}_{\ell t} ) +\phi_{k} u_{\ell t} \} $, the unobserved
element has become a scalar. This seems at first to be an enormous and
disappointing loss of generality, but upon reflection one sees that the loss
is not great. Suppose $u_{\ell t}$ did record a multivariate history over
time, and consider the hazard model $\exp \{ \kappa_{k} (
\overline{Z}_{\ell t},\break  \mathbf{x}_{\ell t} ) +\phi_{k} f (
u_{\ell t} ) \} $ where $f ( \cdot ) $ is some unknown
real-valued functional of that multivariate, temporal history. Although
this appears at first to be a more general model, writing $\widetilde{u}
_{i\ell}=f ( u_{\ell t} )$, the model becomes $\exp \{ \kappa
_{k} ( \overline{Z}_{\ell t}, \mathbf{x}_{\ell t} ) +\phi_{k} %
\widetilde{u}_{i\ell} \} $, a scalar model essentially as before. In
words, in $\exp \{ \kappa_{k} ( \overline{Z}_{\ell t}, \mathbf{x}%
_{\ell t} ) +\phi_{k} f ( u_{\ell t} ) \} $, not
knowing $u_{\ell t}$ and not knowing $f ( \cdot ) $ is no better
and no worse than not knowing the scalar $\widetilde{u}_{i\ell}=f (
u_{\ell t} ) $. It is the impact of unmeasured history on the hazard---a
scalar---that matters, not the particulars of that history. See Li, Propert and
Rosenbaum (\citeyear{LiProRos01}) and \citet{Lu05} for related discussion.

Let $s\in \{ s_{1},\ldots,s_{K} \} $ be one of the point states
or birth outcomes (single girl, etc.), and let $s^{\prime}\neq s$ be any
one of the other states, $s^{\prime}\in$ $ \{ s_{0}, s_{1},\ldots
,s_{K} \} $. Here, $s^{\prime}$ may be either the state $s_{0}$ of
not giving birth or a point state. Suppose that we form a risk-set match
of one individual with $Z_{\ell t}=s$ and $J-1\geq1$ other individuals
$%
\ell^{\prime}$ in state $s^{\prime}$ at $t$, where all $J$ individuals
have the same observed history to time $t$, $\overline{Z}_{\ell
t}=\overline{%
Z}_{\ell^{\prime}t}$ and $\mathbf{x}_{\ell t}=\mathbf{x}_{\ell
^{\prime
}t} $. For instance, this might be a match of $J$ women with the same
observed history to time $t$, one of whom gave birth to her first child
at $%
t $, a single girl $s_{1}$, where the other $J-1$ women had had no
child up
to and including time $t$. Despite looking similar prior to time $t$, it
is possible, perhaps likely, that these $J$ women differed in their
ambitions $u_{\ell t}$ for school or work. After all, one had a child at
time $t$ while the others did not. Alternatively, the matching might
compare a woman who had her first child, a girl or point state $s_{1}$, at
time $t$ to $J-1$ women with the same observable past who had a first child,
a boy or point state $s_{2}$, at time $t$. Perhaps this second comparison
is closer to random than the previous comparison of women with and without
children at time $t$, because now all $J$ women had their first child at
time $t$, and it was only the sex of the child that varied. Obviously,
there are many analogous possibilities, but we suppose the investigator will
focus on one such comparison at a time, for now, $s$ and $s^{\prime}$
with $%
s\neq s^{\prime}$ and $s$, $s^{\prime}\in \{ s_{0},\ldots
,s_{K} \} $.

The risk-set match is built rolling forward in time $t$, matching women with
states $s$ or $s^{\prime}$ at $t$ and with identical observable pasts,
$%
( \overline{Z}_{\ell t}, \mathbf{x}_{\ell t} ) $, possibly
different unobservable pasts $u_{\ell t}$, removing individuals once
matched; however, events subsequent to time $t$ are not used in
matching at
time $t$. In the end, there are $I$ nonoverlapping matched sets, each
containing $J$ individuals. It is notationally convenient to replace the
label $\ell$, where $\ell$ does not indicate who is matched to whom, by
noninformative labels for sets, $i=1,\ldots,I$, and for individuals within
sets, $j=1,\ldots,J$, for instance, random labels could be used. We then
have $\overline{Z}_{ijt}=\overline{Z}_{ij^{\prime}t}$ and $\mathbf{x}%
_{ijt}=\mathbf{x}_{ij^{\prime}t}$ for all $i$, $j$, $j^{\prime}$, but
possibly $u_{ijt}\neq u_{ij^{\prime}t}$. Also, write $\mathcal{F}%
_{it}= ( \overline{Z}_{i1t}, \mathbf{x}_{i1t}, u_{i1t},\ldots,%
\overline{Z}_{iJt}, \mathbf{x}_{iJt}, u_{iJt} ) $. Let $\mathcal{Z}$
be the event that for each $i$, exactly one individual $j$ has $Z_{ijt}=s$
and the remaining $J-1$ individuals $j^{\prime}$ have $Z_{ij^{\prime
}t}=s^{\prime}$, so the risk-set matched design ensures that $\mathcal{Z}$
occurs. Given $\mathcal{Z}$, the time $t$ is fixed, and the two states, $s$
and $s^{\prime}$, are fixed, so it is convenient to write $Z_{ij}=1$
if $%
Z_{ijt}=s$ and $Z_{ij}=0$ if $Z_{ijt}=s^{\prime}$, so that $%
1=\sum_{j=1}^{J}Z_{ij}$ for each $i$.

The next step is key. Although there are ${K+1\choose2}$ possible choices
of two states~$s$, $s^{\prime}\in \{ s_{0},\ldots,s_{K} \} $ to
compare by risk-set matching, the same unobserved covariate $u_{ijt}$ can
severely bias some choices of two states, while others may be nearly random
or only slightly biased. Consider the conditional probability that, in set
$i$ of this risk-set matched design, it is individual $j$ who received
treatment $s$, with $Z_{ijt}=s$, the remaining $J-1$ individuals receiving
treatment $s^{\prime}$. Using (i) $\lambda_{k} ( \overline{Z}%
_{ijt}, \mathbf{x}_{ijt}, u_{ijt} ) =\exp \{ \kappa_{k} (
\overline{Z}_{ijt}, \mathbf{x}_{ijt} ) +\phi_{k} u_{ijt} \} $,
(ii) $\overline{Z}_{ijt}=\overline{Z}_{ij^{\prime}t}$ and $\mathbf
{x}_{ijt}=%
\mathbf{x}_{ij^{\prime}t}$, and (iii) $\sum_{j^{\prime}\neq j}\phi
_{s^{\prime}} u_{ij^{\prime}t}=-\phi_{s^{\prime
}} u_{ijt}+\sum_{j^{\prime}=1}^{J}\phi_{s^{\prime}} u_{ij^{\prime}t}$
yields
{\fontsize{9.8}{11.8}{\selectfont
\begin{eqnarray}\label{eqSenMod}
&&\Pr( Z_{ijt}=s \vert\mathcal{F}_{it},
\mathcal{Z}\nonumber
)
\\
&&\hspace*{-4pt}\qquad=\frac{\exp \{ \kappa_{s} ( \overline{Z}_{ijt}, \mathbf{x}%
_{ijt} ) +\phi_{s} u_{ijt} \} \prod_{j^{\prime}\neq
j}^{J}\exp \{ \kappa_{s^{\prime}} ( \overline{Z}_{ij^{\prime
}t}, \mathbf{x}_{ij^{\prime}t} ) +\phi_{s^{\prime}} u_{ij^{\prime
}t} \} }{\sum_{m=1}^{J}\exp \{ \kappa_{s} ( \overline{Z}%
_{imt}, \mathbf{x}_{imt} ) +\phi_{s} u_{imt} \}
\prod_{m^{\prime}\neq m}^{J}\exp \{ \kappa_{s^{\prime
}} ( \overline{Z}_{im^{\prime}t}, \mathbf{x}_{im^{\prime}t} )
+\phi_{s^{\prime}} u_{im^{\prime}t} \} }\nonumber
\\
&&\hspace*{-4pt}\qquad=\frac{\exp ( \phi_{s} u_{ijt}+\sum_{j^{\prime}\neq j}\phi
_{s^{\prime}} u_{ij^{\prime}t} ) }{\sum_{m=1}^{J}\exp ( \phi
_{s} u_{imt}+\sum_{m^{\prime}\neq m}\phi_{s^{\prime}} u_{im^{\prime
}t} ) }\\
&&\hspace*{-4pt}\qquad=\frac{\exp \{ ( \phi_{s}-\phi_{s^{\prime}} )
u_{ijt} \} }{\sum_{m=1}^{J}\exp \{ ( \phi_{s}-\phi
_{s^{\prime}} ) u_{imt} \} }\nonumber
\\
&&\hspace*{-4pt}\qquad=\frac{\exp ( \gamma u_{ijt} ) }{\sum_{m=1}^{J}\exp (
\gamma u_{imt} ) }\qquad\mbox{where }\gamma=\phi_{s}-\phi
_{s^{\prime}}, \nonumber
\end{eqnarray}}}
\hspace*{-3pt}where the last expression (\ref{eqSenMod}) is the same as the sensitivity
analysis model in Rosenbaum (\citeyear{Ros07,Ros13N2}) for
comparing treatment and
control in $I$ matched sets.

The key point is that there may be reason to believe that $\llvert \phi
_{s}-\phi_{s^{\prime}}\rrvert $ is small for some choices of $s$, $%
s^{\prime}$, and large for other choices. Refraining from having a child,
$s=0$, is often a carefully planned event, but whether a child is a boy
or a
girl, twins or a single birth, is a much more haphazard event. Some
comparisons are expected to be less biased by unmeasured intentions and
preferences than other comparisons. If a careful choice of $s$,
$s^{\prime
}$ implies that $\llvert \gamma\rrvert =\llvert \phi_{s}-\phi
_{s^{\prime}}\rrvert $ is small, then the inference about treatment
effects may be convincing if it is insensitive to small biases $\llvert
\gamma\rrvert $ even if it is sensitive to moderate biases. If $\phi
_{s}-\phi_{s^{\prime}}=0$, then (\ref{eqSenMod}) is the randomization
distribution, $\Pr ( Z_{ijt}=s \vert \mathcal{F}_{it}, %
\mathcal{Z} ) =1/J$ for each $ijt$; moreover, this is true even if $%
\phi_{s}$ and $\phi_{s^{\prime}}$ are large, so that comparing mothers
who had children at different times would be severely biased by $u_{ijt}$.

%s2.4 #&#
\subsection{\texorpdfstring{Sensitivity analysis for any remaining differential biases.}
{Sensitivity analysis for any remaining differential biases}}
\label{ssSenMod}

If $\phi_{s}\neq\phi_{s^{\prime}}$, but $\llvert \gamma\rrvert
=\llvert \phi_{s}-\phi_{s^{\prime}}\rrvert $ is small in (\ref%
{eqSenMod}), then the differential comparison of treatments $s$ and $%
s^{\prime}$ in (\ref{eqSenMod}) may still be biased by $u_{ijt}$, and the
sensitivity analysis examines the possible consequences of biases of various
magnitudes $\gamma$. In the current paper, the sensitivity analyses
use (%
\ref{eqSenMod}) with a test statistic that is either the mean
difference in
workforce participation or a corresponding $M$-estimate with Huber's
weights. Of course, the mean difference is one particular form of $M$%
-estimate. The sensitivity analysis was implemented as described in
\citet{Ros07} with the restriction that $u_{ijt}\in [ 0,1 ] $,
so that under (\ref{eqSenMod}) matched mothers may differ in their hazards
of birth outcome $s$ versus $s^{\prime}$ by at most a factor of $\Gamma
=\exp ( \gamma ) $. In the comparison in Section~\ref{secAnalysis},
this means that two mothers with the same pattern of fertility and observed
covariates to time $t$, both of whom gave birth at time $t$, may differ in
their odds of having a twin, $s$, rather than a single child of a different
sex than her earlier children, $s^{\prime}$, by at most a factor of
$\Gamma
$ because of differences in the unmeasured~$u_{ij}$. Although biases of
this sort are not inconceivable, perhaps as a consequence of differential
use of abortion or fertility treatments, presumably such a bias $\Gamma
$ is
not very large, much smaller than the biases associated with efforts to
control the timing of births. The one parameter $\Gamma$ may be
reinterpreted in terms of two parameters describing treatment-control pairs,
one $\Delta$ relating $u_{ij}$ to the outcome $ (
r_{Tij},r_{Cij} ) $, the other $\Lambda$ relating $u_{ij}$ to the
treatment $Z_{ij}$, such that a single value of $\Gamma$ corresponds
to a
curve of values of $ ( \Delta,\Lambda ) $ defined by $\Gamma
= ( \Delta\Lambda+1 ) / ( \Delta+\Lambda ) $, so a
brief unidimensional analysis in terms of $\Gamma$ may be interpreted in
terms of infinitely many two-dimensional analyses in terms of $ ( \Delta
,\Lambda ) $; see \citet{RosSil09}. For instance, the
curve for $\Gamma= ( \Delta\Lambda+1 ) / ( \Delta+\Lambda
) =1.25$ includes the point $ ( \Delta,\Lambda ) = (
2,2 ) $ for a doubling of the odds of treatment and a doubling of the
odds of a positive pair difference in outcomes. \citet{HsuSma13} show
how to calibrate a sensitivity analysis about an unobserved covariate using
the observed covariates.

What is the role of the restriction $u_{ijt}\in [ 0,1 ] $? The
restriction $u_{ijt}\in [ 0,1 ] $ gives a simple numerical meaning
to $\gamma$ and $\Gamma$ by fixing the scale of the unobserved covariate:
in (\ref{eqSenMod}), two subjects may differ in their hazard of
treatment $s$
rather that treatment $s^{\prime}$ at time $t$ by at most a factor of $
\Gamma$ because they differ in terms of $u_{ijt}$. It is possible to
replace the restriction that $u_{ijt}\in [ 0,1 ] $ for all $ijt$
by the restriction that $u_{ijt}\in [ 0,1 ] $ for, say, 99\% of
the $ijt$ with the remainder unrestricted [Rosenbaum (\citeyear{Ro87}),
Section~4]; however,
when using robust methods, small parts of the data make small contributions
to the inference, so this replacement has limited impact. Permitting 1\%
of the $u_{ijt}$ to be unrestricted should count as a larger bias, in some
sense a larger $\gamma$, and \citet{WanKri06} explore this
possibility in a special case, concluding that binary $u_{ijt}$ do the most
damage among all $u_{ijt}$ with a fixed standard deviation.

For discussion of a variety of methods of sensitivity analysis in
observational studies, see \citet{Baietal10}, \citet{CORetal59},
\citet{DipGan04}, Egleston, Scharfstein and
MacKenzie (\citeyear{EglSchMac09}), \citet{Gas92}, Hosman, Hansen and Holland
(\citeyear{HosHanHol10}), Li, Propert and Rosenbaum (\citeyear{LiProRos01}), Lin, Psaty and Kronmal (\citeyear{LinPsaKro98}),
Liu, Kuramoto and Stuart (\citeyear{LiuKurStu13}),
\citet{Mar97}, McCandless, Gustafson and
Levy (\citeyear{McCGusLev07}), Robins, Rotnitzky and
Scharfstein (\citeyear{RobRotSch00}),
Rosenbaum (\citeyear{Ros07,Ros13N2}), \citet{Sma07}, \citet{SmaRos08} and
\citet{YuGas05}.

%s2.5 #&#
\subsection{What is isolation?}
\label{ssWhatIsIsolation}

Isolation refers to equation (\ref{eqSenMod}) and is motivated by the
possibility that $\llvert \phi_{s}-\phi_{s^{\prime}}\rrvert $ may
be small or zero when neither $\phi_{s}$ nor $\phi_{s^{\prime}}$ is small
or zero. If $\phi_{s}$ is not small, receipt of treatment $s$ rather than
no treatment will be biased by the unmeasured time-dependent covariate $
u_{ijt}$. In parallel, if $\phi_{s^{\prime}}$ is not small, receipt of
treatment $s^{\prime}$ rather than no treatment will be biased by
$u_{ijt}$%
. However, if $\phi_{s}=\phi_{s^{\prime}}$, then the differential
comparison of treatments $s$ and $s^{\prime}$, conditionally given one of
them, will not be biased by $u_{ijt}$, even though $\phi_{s}$ and $\phi
_{s^{\prime}}$ may both be large. If unmeasured aspirations and plans
($%
u_{ijt}$) influence the timing of fertility but not whether twins ($s$)
or a
single child ($s^{\prime}$) is born, then a comparison of two mothers with
the same timing, one with twins, the other with a single child, is not
biased by the unmeasured aspirations and plans. Equation (\ref{eqSenMod})
isolates biased timing from possibly unbiased birth outcomes given
timing. The sensitivity analysis considers the possibility that
$\llvert \phi
_{s}-\phi_{s^{\prime}}\rrvert $ is small but not zero, so there is
some differential bias.

In the case study, it seems likely that the timing of births is
affected by
unmeasured covariates $u_{ijt}$ but, conditionally given a birth, specific
birth outcomes are close to random; that is, each $\phi_{s}$ is not small
but each $\llvert \phi_{s}-\phi_{s^{\prime}}\rrvert $ is small. In
some other context, it might be that $\llvert \phi_{s}-\phi
_{s^{\prime
}}\rrvert $ is thought to be small for some pairs $s$, $s^{\prime}\in
\{ 1,\ldots,K \} $ and not for others, and, in this case, attention
might be restricted to a few comparisons for which $\llvert \phi
_{s}-\phi_{s^{\prime}}\rrvert $ is thought to be small.

No matter how deliberate and purposeful a life may be, there are brief
moments when some consequential aspect of that life is determined by
something haphazard. Isolation narrows the focus in two ways: the
moment and
the aspect. One compares people who appeared similar a moment before luck
played its consequential role. Among such people, one considers only a
consequential aspect controlled by luck. Isolation refers to the joint use
of risk-set matching to focus on a moment and differential effects to focus
on an aspect.

%s2.6 #&#
\subsection{\texorpdfstring{Selecting strong but haphazard comparisons.}{Selecting strong but haphazard comparisons}}
\label{ssStrongButHaphazard}

To emulate a randomized experiment, a natural experiment should have a
consequential difference in treatments determined by something
haphazard. The strongest contrast is twins at birth $k$ versus mixed
sex children at
birth $k$, because this comparison is expected to do the most to shift the
number of children. The population of pregnant women would not be
distorted by limiting attention to these two groups, providing that the
unobserved $u_{ijt}$ affects the timing but not the outcome of pregnancies
(i.e., providing $\phi_{s}=\phi_{s^{\prime}}$ for $s$, $s^{\prime
}\in
\{ 1,\ldots,K \} $).

Natural experiments may yield instrumental variables where
``strong'' refers to the strength of the instrument. An
instrument is a haphazard nudge to accept a higher dose of treatment, where
the nudge affects the outcome only if it alters the dose of treatment, the
so-called ``exclusion restriction''; see
\citet{Hol88} and Angrist, Imbens and Rubin (\citeyear{AngImbRub96}). In Section~\ref{ssDifRiskMatch},
some patterns of births (e.g., twins) may induce women to have more children
than they would have had with a different pattern of births, so perhaps
certain patterns are instruments for family size (the dose). An instrument
is weak if most nudges are ignored, rarely altering the dose. An
instrument is strong if it typically materially alters the dose. Weak
instruments create inferential problems with limited identification [Bound, Jaeger and Baker
(\citeyear{BouJaeBak95}), Imbens and Rosenbaum (\citeyear{ImbRos05}), \citet{Sma07}] and,
more importantly,
inferences based on weak instruments are invariably sensitive to tiny
departures from randomized assignment\vadjust{\goodbreak} [\citet{SmaRos08}].
Therefore, it is often advantageous to strengthen an instrument
[\citet{Baietal10}, Zubizarreta et al. (\citeyear{Zubetal13})].

Is the exclusion restriction plausible here? Perhaps not. The exclusion
restriction would mean that having twins affects workforce participation
only by altering the total number of children. If a mother wanted three
children but had twins at her second pregnancy, the occurrence of twins
might have altered the timing of her children's births rather than the total
number of children. A mother who wished to stay at home until her three
children had entered kindergarten might return to work sooner because of
twins at the second birth without altering her total number of
children, and
in this case the exclusion restriction would not be satisfied.

Even if the exclusion restriction does not hold, so the natural experiment
does not yield an instrument, it is nonetheless advantageous to have a
consequential difference in treatments determined by something that is
haphazard. In particular, the Wald estimator commonly used with
instrumental variables estimates a ratio of treatment effects---a
so-called effect ratio---when the exclusion restriction does not
hold. The effect ratio is a local-average treatment effect when the exclusion
restriction holds, but it is interpretable without that condition; see
Section~\ref{secAnalysis} and \citet{Baietal10} for further discussion.

A distinction is sometimes made between internal and external validity, a
distinction introduced by Donald T. Campbell and colleagues, a distinction
that \citet{Cam86} later attempted to revise. In revised form, internal
validity became ``local causal
validity,'' meaning correct estimation of the effects
of the treatments actually studied
in the populations actually studied. What had been external validity
separated into several concepts, each referring to some generalization,
perhaps from the treatments under study to other related treatments, from
the populations under study to other related populations, or from the
outcome measures under study to other related measures. Because it uses
Census data from 1980, Angrist and Evans' (\citeyear{AngEva96}) study concerns of a
well-defined population at a particular era in history, and results about
women's workforce participation might easily be different in the US in
earlier and later eras. It would be comparatively straightforward to
replicate their study using Census data from other eras or using similar
data in other countries. Their study is reasonably compelling as a study
of the effects of having twins rather than a single child but, as the
discussion of the exclusion restriction above makes clear, it is not certain
that having twins has the same effect on workforce participation as having
two children at different times. Moreover, the study provides no
information about women who have no children at all. In brief, twinning is
typically an unintended and somewhat random event, whereas many women
attempt to carefully, thoughtfully and deliberately control the timing of
fertility, so Angrist and Evan's study has unusual strengths in local causal
validity, but one needs to avoid extrapolating their findings to other eras
or types of fertility that they did not study.\vadjust{\goodbreak}

%s3 #&#
\section{\texorpdfstring{The risk-set match.}{The risk-set match}}
\label{secAEconstruction}

%s3.1 #&#
\subsection{\texorpdfstring{One matched risk set.}{One matched risk set}}
\label{ssOneMatchedRiskSet}

We created nonoverlapping matched sets of 6 women who were similar
prior to
the birth of their $k$th child, for $k=2$, 3, 4, one of whom had a twin on
this $k$th birth, whereas the others had children of both sexes as of
the $k$th child. For instance, matched set \#836 consisted of six
women. All
six women had their first child at age 18 and their second child at age 22,
and all were white. After the birth of the second child, five of the
mothers had one boy and one girl, and one of the mothers had twins at the
second pregnancy. A~mother's plans for education, career and family may
easily influence the timing of her pregnancies, but these six women gave
birth at the same ages. A mother's plans for education, career and family
are much less likely to determine which of the six pregnancies will end with
twins and which will end with two children of different sexes---for most
mothers, that's just luck. All six mothers had 12 years of education at
the time of their first and second births at ages 18 and 22, respectively;
see Section~\ref{sec3.2} for technical details about this statement.

Matched sampling controls, or should control, for the past, not the future
[\citet{Ros84}]. The six women were similar prior to their second
pregnancy. They had different outcomes at their second pregnancy. What
happened subsequently? The woman with twins ended up with 3 children in
total, the other five woman ended up with two children each---that is,
none of these women went on to have additional children beyond their second
pregnancy. The pattern is different in other matched sets. In this one
matched set, all six women had no additional education beyond the 12 years
they had at age 18, the age of their first birth. In this particular
matched set, the mother of twins ranked third in workforce
participation. In the year prior to the 1980 Census, two of the women
with two children had
worked at least 40 hours in the previous week and 52 weeks in the previous
year, while the remaining three women with two children had not worked at
all in the previous year. The woman with twins, with three children, had
worked 40 hours in the previous week and 20 weeks in the previous year.

Matched sets varied, but set \#836 was typical in one respect. In the
matched comparison, it was uncommon for women who had children by age
18 to
ultimately complete a BA degree---only 5.5\% did so---whereas it was
much more common for women who did not have a child by age 18 to
complete a
BA degree---28.2\% did so. Total lifetime education is the sum of two
variables, a covariate describing education prior to the $k$th birth
and an
outcome describing additional education subsequent to the $k$th birth.
Risk-set matching entails matching for the covariate---the past---but
not for the outcome---the future.

%s3.2 #&#
\subsection{\texorpdfstring{Technical detail: How the matching was done.}
{Technical detail: How the matching was done}}\label{sec3.2}

Matches were constructed in temporal order, beginning with the second
pregnancy. Mothers not matched at the second pregnancy might be matched
later. The matching was exact for three variables---age category at the
second pregnancy, race/ethnicity and region of the US; see Table
%TCIMACRO{\TeXButton{reftabCov1}{\ref{tabCov1}}}%
%BeginExpansion
\ref{tabCov1}%
%EndExpansion
. Within each of these $64=4^{3}$ cells, the match solved a combinatorial
optimization problem to make the mother of twins similar to the five control
mothers in the same matched set. Similarity was judged by a robust
Mahalanobis distance [\citet{Ros10}, Section~8.3] using observed
covariates $%
\mathbf{x}_{it}$ prior to this pregnancy. Forming nonoverlapping matched
sets to minimize the sum of the treated-versus-control distances within sets
is a version of the optimal assignment problem, and it may be solved using
the \texttt{pairmatch} function of Hansen's (\citeyear{Han07}) \texttt
{optmatch} package
in \texttt{R}. [We used \texttt{mipmatch} in \texttt{R} available at
\surl{http://www-stat.wharton.upenn.edu/\textasciitilde josezubi/}; see Zubizarreta (\citeyear{Zub12}).]

From the Census data, we can know the education of the mother prior to the
Census, her age at the Census and the ages of her children, and from
this we
can deduce her ages at the births of her children. Ideally, we would know
exactly her years of education at the birth of each of her children,
but the
Census provides slightly less information. The norm in the US is to
complete high school with 12 years of education at age 18. If a woman had
a total of $E$ years of education at the time of the census and if she was
age $A$ at her $k$th pregnancy, we credited her with $\min (
E,A-6 ) $ years of education at her $k$th pregnancy. For instance, a~woman who had a BA degree with 16 years of education and a first child at
age 26 was credited with 16 years of education at the birth of her first
child. This is a reasonable approximation but will err in some cases.
The exact timing of education is available in some longitudinal data sets.

%s3.3 #&#
\subsection{\texorpdfstring{Covariate balance prior to the $k$th birth in the risk-set match.}{Covariate balance prior to the $k$th birth in the risk-set match}}
\label{ssBalance}

\begin{figure}

\includegraphics{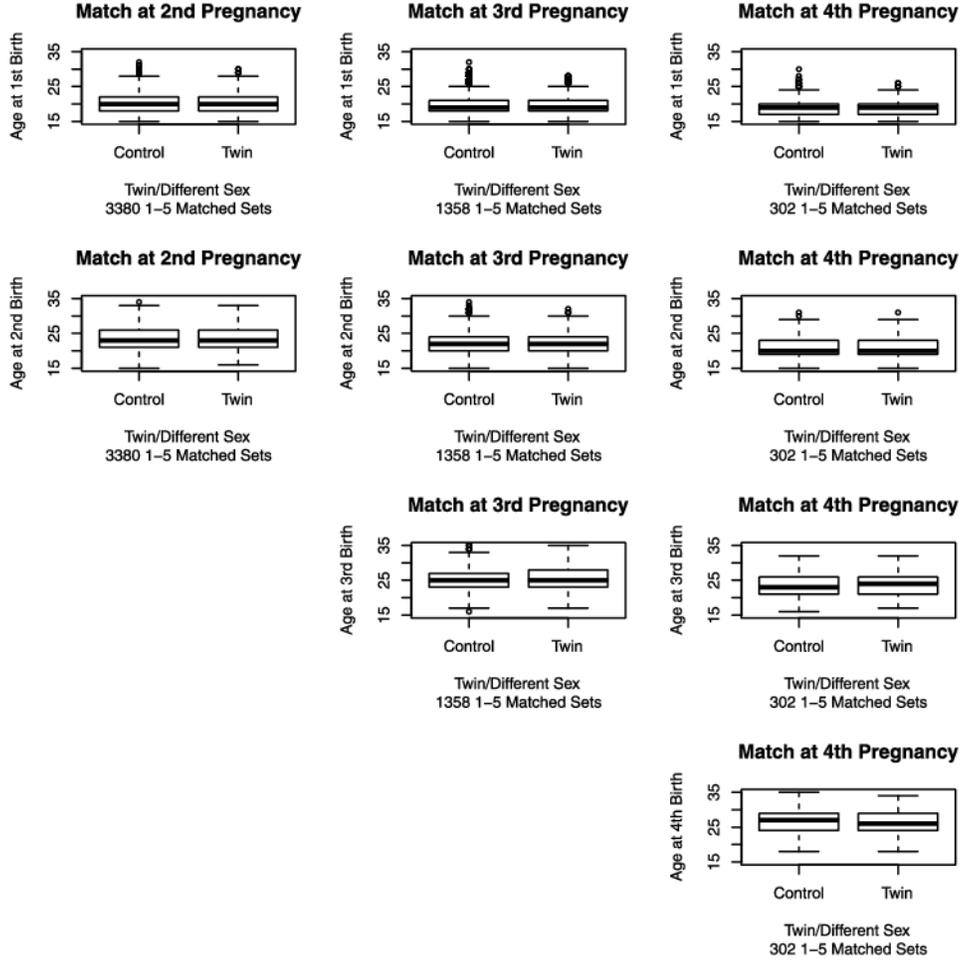}

  \caption{Age at births in 5040 1--5 nonoverlapping matched sets containing
30,240 mothers, specifically 5040 mothers who gave birth to a twin at the
indicated pregnancy and 25,200 mothers who had at least one child of each
sex by the end of the indicated pregnancy. For 3380 sets matched at the
second pregnancy, matching controlled the past, namely age at the first and
second births. For 1358 sets matched at the third pregnancy, matching
controlled the past, namely age at the first, second and third births. For
302 sets matched at the fourth pregnancy, matching controlled the past,
namely age at the first, second, third and fourth births.}\label{fig1}
\end{figure}

\begin{figure}

\includegraphics{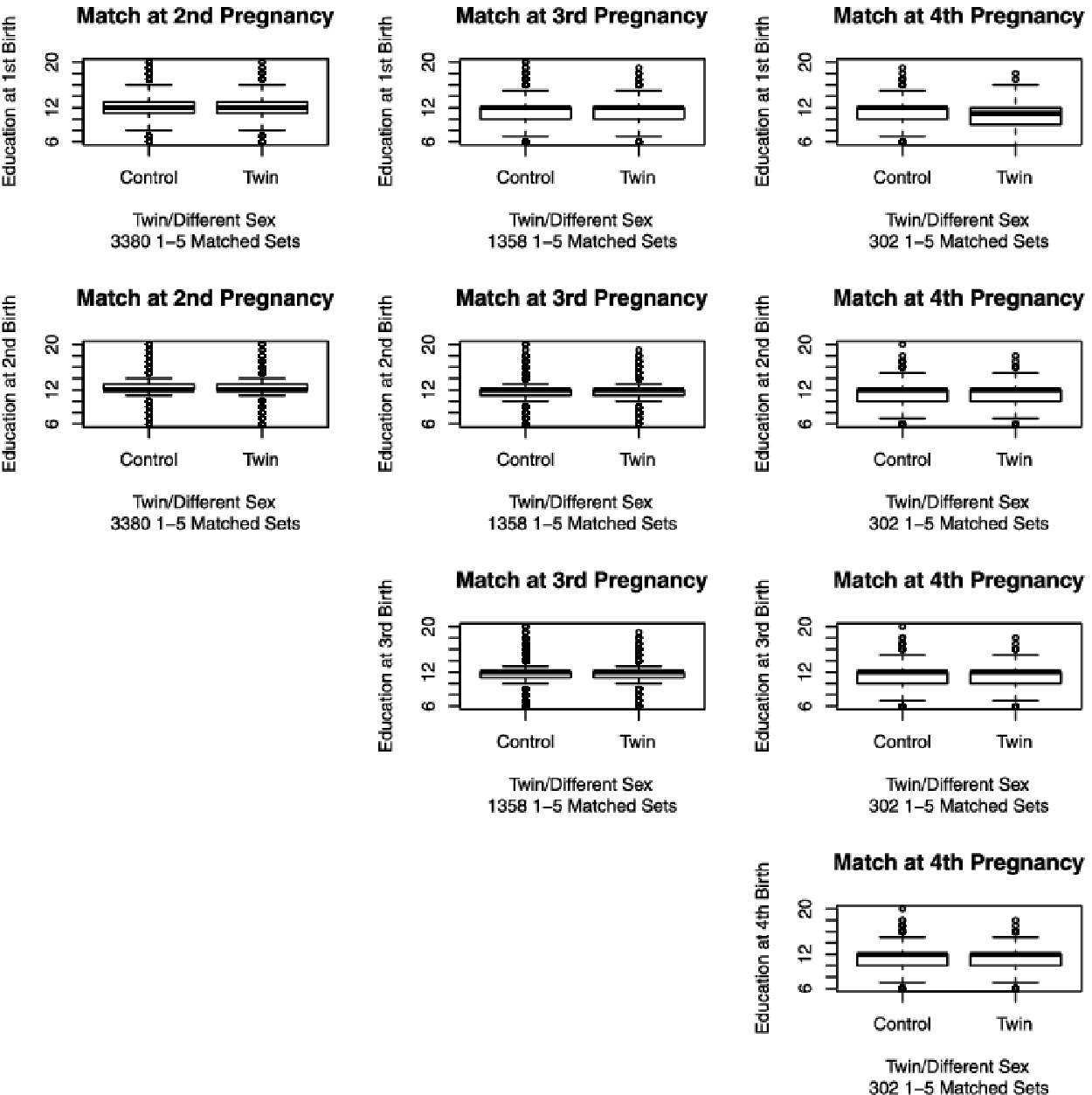}

  \caption{Mother's education at the time of various births in 5040 1--5
nonoverlapping matched sets containing 30,240 mothers, specifically 5040
mothers who gave birth to a twin at the indicated pregnancy and 25,200
mothers who had at least one child of each sex by at the end of the
indicated pregnancy. Each match controls the past, not the future. For
graphical display in the boxplots, education is truncated at 6 years
despite a few values below that.}\label{fig2}
\end{figure}

Figures~\ref{fig1} and \ref{fig2} show the balance on age at each pregnancy and education at
each pregnancy. The match at the second pregnancy should balance age and
education at the first two pregnancies, viewing subsequent events as
outcomes. The match at the third pregnancy should balance age and
education at the first three pregnancies, viewing subsequent events as
outcomes. The match at the fourth pregnancy is analogous. Figures~\ref{fig1} and
\ref{fig2} show the desired balance was achieved.

Tables
%TCIMACRO{\TeXButton{reftabCov1}{\ref{tabCov1}} }%
%BeginExpansion
\ref{tabCov1}
%EndExpansion
and
%TCIMACRO{\TeXButton{reftabCov2}{\ref{tabCov2}} }%
%BeginExpansion
\ref{tabCov2}
%EndExpansion
show the comparability of the matched groups separately for the matches at
the second, third and fourth pregnancy. Table
%TCIMACRO{\TeXButton{reftabCov1}{\ref{tabCov1}} }%
%BeginExpansion
\ref{tabCov1}
%EndExpansion
exhibits perfect balance for categories of race/ethnicity, region of
the US
and age at the second pregnancy. Moreover, the interactions of these three
variables are also exactly balanced.

%t1 #&#
\begin{table}
\tabcolsep=0pt
\caption{In each matched risk set containing $J=6$ mothers, a mother of
a twin at birth $k$ is matched to $J-1=5$
control mothers whose $k$th birth was a single child whose sex was
different from one of her previous children. The
matching was exact for four age categories, for four race/ethnicity
categories and for four regions of the US,
and because it was exact, it controlled their interactions. The table
displays counts and percents, where the count for controls is
always five times the count for twins. Only one column of percents is
displayed because the percents in the two groups
are identical}
\label{tabCov1}\label{balance1}
\begin{tabular*}{\textwidth}{@{\extracolsep{4in minus 4in}}ld{4.0}d{5.0}d{2.0}d{4.0}d{4.0}d{2.0}d{3.0}d{4.0}c@{}}
\hline
\multicolumn{1}{c}{} & \multicolumn{3}{c}{\textbf{2nd birth}} & \multicolumn
{3}{c}{\textbf{3rd birth}} & \multicolumn{3}{c@{}}{\textbf{4th birth}} \\[-6pt]
\multicolumn{1}{c}{} & \multicolumn{3}{c}{\hrulefill} & \multicolumn
{3}{c}{\hrulefill} & \multicolumn{3}{c@{}}{\hrulefill} \\
\multicolumn{1}{@{}l}{\textbf{Covariate}} &
\multicolumn{1}{c}{\textbf{Twin}} & \multicolumn{1}{c}{\textbf{Control}} & \multicolumn{1}{c}{\textbf{\%}} &
\multicolumn{1}{c}{\textbf{Twin}} & \multicolumn{1}{c}{\textbf{Control}} & \multicolumn{1}{c}{\textbf{\%}} &
\multicolumn{1}{c}{\textbf{Twin}} & \multicolumn{1}{c}{\textbf{Control}}
& \multicolumn{1}{c@{}}{\textbf{\%}} \\
\hline
Age & \multicolumn{9}{c}{Mother's age at her second pregnancy} \\
$\le18$ &182 & 910 & 5 & 167 & 835 & 12 & 63 & 315 & 21\\
19--22 & 1239 & 6195 & 37 & 677 & 3385 & 50 & 163 & 815 & 54\\
23--25 & 1044 & 5220 & 31 & 350 & 1750 & 26 & 63 & 315 & 21\\
$\ge26$ & 915 & 4575 & 27 & 164 & 820 & 12 & 13 & 65 & \phantom{0}4\\ [3pt]
Race/ethnicity & \multicolumn{9}{c}{Mother's race/ethnicity} \\
Black & 505 & 2525 & 15 & 242 & 1210 & 18 & 81 & 405 & 27\\
Hispanic & 87 & 435 & 3 & 63 & 315 & 5 & 11 & 55 & \phantom{0}4\\
White & 2707 & 13\mbox{,}535 & 80 & 1023 & 5115 & 75 & 203 & 1015 & 67\\
Other & 81 & 405 & 2 & 30 & 150 & 2 & 7 & 35 & \phantom{0}2\\ [3pt]
Region & \multicolumn{9}{c}{Region of the US} \\
Northeast & 685 & 3425 & 20 & 270 & 1350 & 20 & 60 & 300 & 20\\
South & 1081 & 5405 & 32 & 426 & 2130 & 31 & 100 & 500 & 33\\
Central & 988 & 4940 & 29 & 391 & 1955 & 29 & 93 & 465 & 31\\
West & 626 & 3130 & 19 & 271 & 1355 & 20 & 49 & 245 & 16\\
\hline
\end{tabular*}
\end{table}

\begin{table}
\caption{Baseline comparison of 30,240 distinct mothers in
$I=5040=3380+1358+302$ nonoverlapping matched sets
of $J=6$ mothers, each set containing one mother who gave birth to a
twin and $J-1$ control mothers who gave birth
to a single child whose sex differed from that of one of her previous
children. The table shows age and education of
mothers at their various births prior to risk-set matching}
\label{tabCov2}\label{balance2}
\begin{tabular*}{\textwidth}{@{\extracolsep{4in minus 4in}}ld{2.1}d{2.1}d{2.1}d{2.1}d{2.1}d{2.1}@{}}
\hline
\multicolumn{1}{c}{} & \multicolumn{2}{c}{\textbf{2nd birth}} & \multicolumn
{2}{c}{\textbf{3rd birth}} & \multicolumn{2}{c@{}}{\textbf{4th birth}} \\[-6pt]
\multicolumn{1}{c}{} & \multicolumn{2}{c}{\hrulefill} & \multicolumn
{2}{c}{\hrulefill} & \multicolumn{2}{c@{}}{\hrulefill} \\
\multicolumn{1}{@{}l}{\textbf{Covariate}} &
\multicolumn{1}{c}{\textbf{Twin}} & \multicolumn{1}{c}{\textbf{Control}} &
\multicolumn{1}{c}{\textbf{Twin}} & \multicolumn{1}{c}{\textbf{Control}} &
\multicolumn{1}{c}{\textbf{Twin}} & \multicolumn{1}{c@{}}{\textbf{Control}}\\
\hline
& \multicolumn{6}{c}{Sample size} \\
\# of mothers & \multicolumn{1}{c}{3380} & \multicolumn{1}{c}{16,900} & \multicolumn{1}{c}{1358} & \multicolumn{1}{c}{6790} & \multicolumn{1}{c}{302} & \multicolumn{1}{c}{1510} \\ [3pt]
& \multicolumn{6}{c}{Mother's age in years, mean} \\
At the census & 30.4 & 30.4 & 30.7 & 30.7 & 31.6 & 31.6 \\
At 1st birth & 20.4 & 20.4 & 19.5 & 19.5 & 18.8 & 18.8 \\
At 2nd birth & 23.5 & 23.4 & 21.8 & 21.8 & 20.7 & 20.7 \\
At 3rd birth & & & 25.1 & 25.1 & 23.5 & 23.4 \\
At 4th birth & & & & & 26.7 & 26.6 \\[3pt]
& \multicolumn{6}{c}{Mother's education in years, mean} \\
At 1st birth & 11.9 & 12.0 & 11.4 & 11.4 & 10.8 & 10.9 \\
At 2nd birth & 12.2 & 12.2 & 11.6 & 11.6 & 11.0 & 11.1 \\
At 3rd birth & & & 11.6 & 11.6 & 11.1 & 11.2 \\
At 4th birth & & & & & 11.1 & 11.2 \\ [3pt]
& \multicolumn{6}{c}{Mother's education at 1st birth, \%} \\
High school & 43 & 43 & 42 & 42 & 32 & 33 \\
Some college & 19 & 19 & 14 & 14 & 15 & 14 \\
BA or more & 09 & 09 & 05 & 05 & 03 & 03 \\ [3pt]
& \multicolumn{6}{c}{Mother's education at 2nd birth, \%} \\
High school & 47 & 47 & 48 & 48 & 39 & 40 \\
Some college & 20 & 20 & 15 & 15 & 16 & 15 \\
BA or more & 11 & 11 & 06 & 06 & 04 & 04 \\[3pt]
& \multicolumn{6}{c}{Mother's education at 3rd birth, \%} \\
High school & & & 48 & 48 & 41 & 41 \\
Some college & & & 16 & 16 & 16 & 16 \\
BA or more & & & 06 & 06 & 05 & 05 \\ [3pt]
& \multicolumn{6}{c}{Mother's education at 4th birth, \%} \\
High school & & & & & 41 & 41 \\
Some college & & & & & 16 & 16 \\
BA or more & & & & & 05 & 05 \\
\hline
\end{tabular*}
\end{table}
%

%s4 #&#
\section{\texorpdfstring{Inference: Tobit effects, proportional effects, sensitivity
analysis.}{Inference: Tobit effects, proportional effects, sensitivity
analysis}}
\label{secAnalysis}

\begin{figure}

\includegraphics{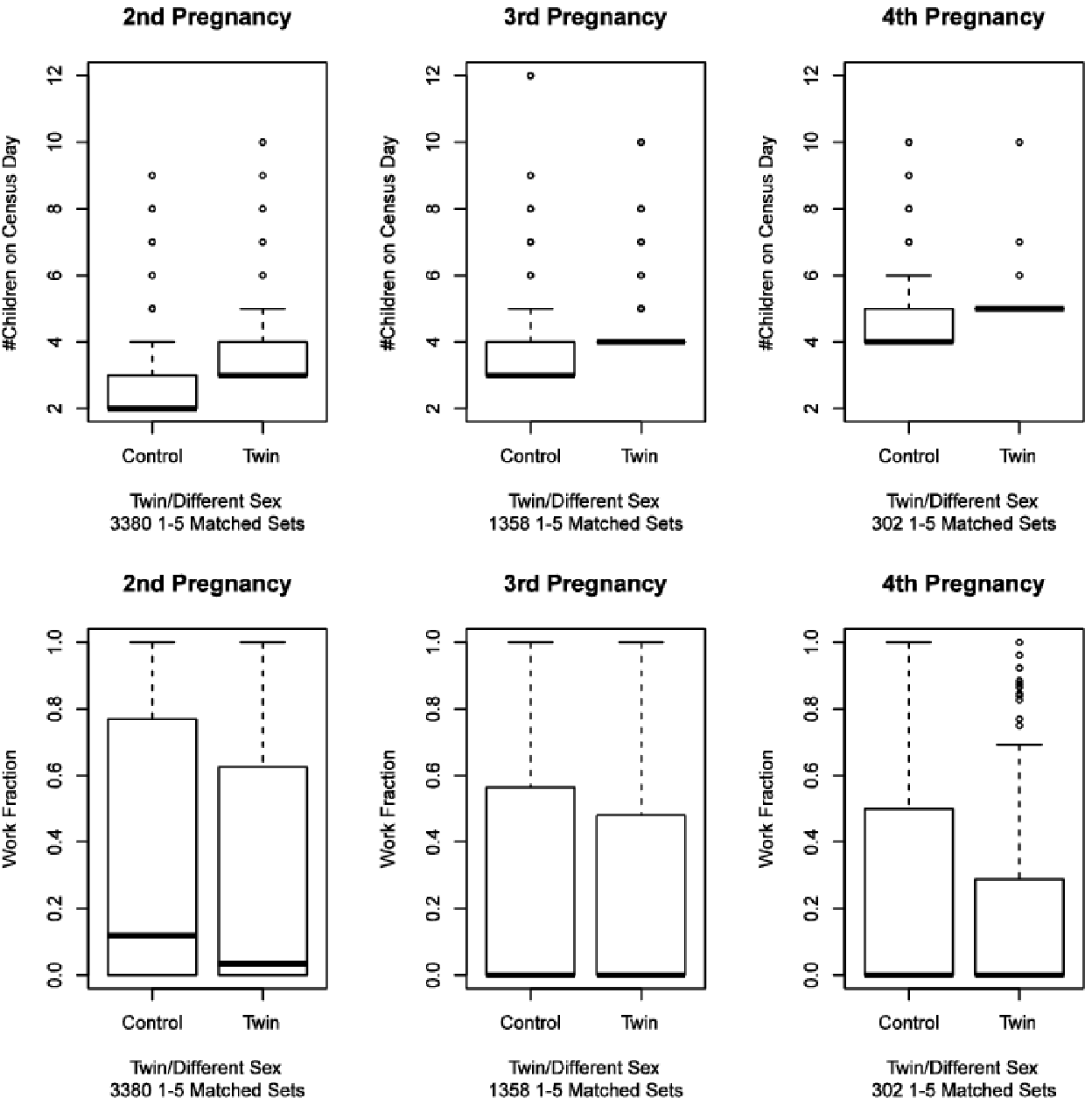}

  \caption{Two outcomes in 5040 1--5 nonoverlapping matched sets containing
30,240 mothers, specifically 5040 mothers who gave birth to a twin at the
indicated pregnancy and 25,200 mothers who had at least one child of each
sex by at the end of the indicated pregnancy. The upper boxplots indicate
the number of children. The lower boxplots indicate the work fraction,
defined to be min(hours worked in the previous week, 40)${}\times{}$(weeks
worked in
the previous year)${}/(40\times 52)$, so a value of 1 is similar to ``full time
employment.''}\label{fig3}
\end{figure}

Figure~\ref{fig3} depicts two outcomes recorded on Census day for the 30,240 mothers
in 5040 matched sets, each set containing one mother who had a twin at the
indicated pregnancy and 5 mothers who had at least one child of each
sex at
the indicated pregnancy. One outcome is the total number of children
recorded on Census day. The other outcome is the work fraction where 0
indicates no work for pay and 1 indicates full time work ($\geq40$ hours
per week). The work fraction is the number of weeks worked in the last
year multiplied by the minimum of 40 and the number of hours worked in the
last week, and then this product is divided by $40\times52$ to produce a
number between 0 and 1. (A small fraction of mothers worked substantially
more than 40 hours in the previous week.)

\begin{figure}

\includegraphics{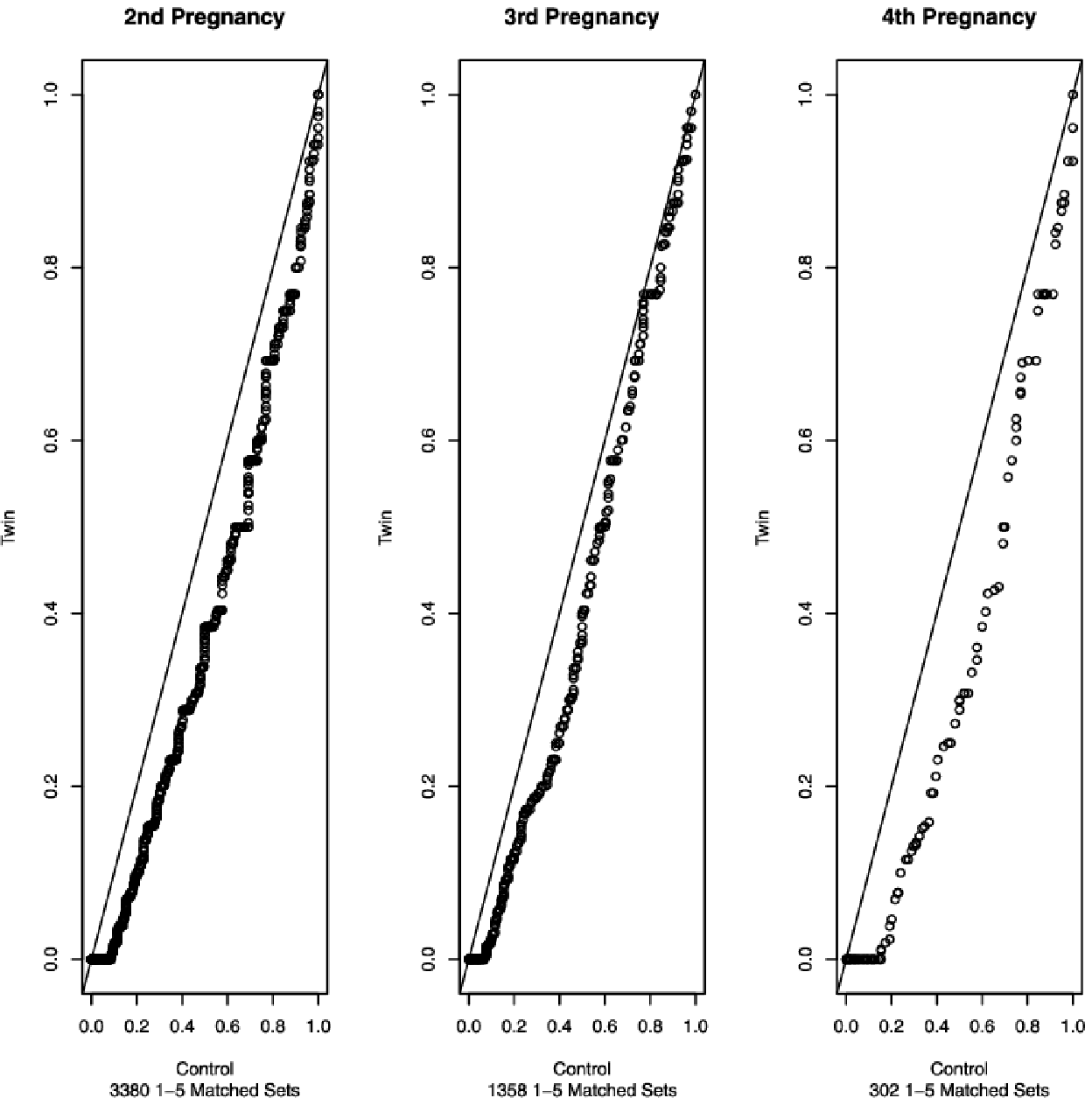}

  \caption{Quantile--quantile plots of work fraction for twins (vertical) and
controls (horizontal) with the line of equality. The plot shows that women
with twins were more likely to not work, as seen in the horizontal
start to
the plot, and they worked fewer hours in total, as quantiles fall below the
line of equality.}\label{fig4}
\end{figure}

In the top half of Figure~\ref{fig3}, at the second pregnancy, a twin birth shifted
upward by about 1 child the boxplot of number of children. The shift is
smaller at the third and fourth pregnancies, where the lower quartile and
median increase by 1 child, but the upper quartile is unchanged.
Presumably, some mothers pregnant for the third or fourth time intend to
have large families and twins did not alter their plans. In the bottom
half of Figure~\ref{fig3}, mothers of twins worked somewhat less, but the difference
in work fraction is not extremely large. Figure~\ref{fig4} displays the information
about work fraction in a different format, as a quantile--quantile plot.

We consider two models for the effect on the fraction worked, $R_{ij}$.
One model is a so-called Tobit effect, named for James Tobin, of twin versus
different-sex-single-child, $Z_{ij}$. The Tobit effect has $r_{Tij}=\max
( 0, r_{Cij}-\tau ) $ and it reflects the fact that a woman's
workforce participation may decline to zero but not further. For instance,
if $\tau=0.1=10\%$, then a mother who would have worked at least $%
r_{Cij}=10\%$ of full-time without twins would work 10\% less with
twins, $%
r_{Tij}= r_{Cij}-10\%$, but a mother who would have worked $r_{Cij}=5\%
$ or
$r_{Cij}=0\%$ of full-time without twins would not work with twins, $%
r_{Tij}=0\%$. For the Tobit effect, we draw inferences about $\tau$. If
$H_{0}\dvtx\tau=\tau_{0}$ were true, then $\max \{ 0, R_{ij}- (
1-Z_{ij} ) \tau_{0} \} =r_{Tij}$ does not vary with $Z_{ij}$
and satisfies the null hypothesis of no treatment. Therefore,
$H_{0}\dvtx\tau
=\tau_{0}$ is the hypothesis of no treatment effect on $\max \{
0, R_{ij}- ( 1-Z_{ij} ) \tau_{0} \} $ and the confidence
interval is obtained in the usual way by inverting the test. In the usual
way, the point estimate solves for $\tau$ an estimating equation that
equates the test statistic to its null expectation. We use the
treated-minus-control mean as the test statistic, but very similar results
were obtained using an $M$-estimate with Huber's weight function
trimming at
twice the median absolute deviation. See \citet{Ros07} and the
\texttt{%
senmwCI} function in the \texttt{sensitivitymw} package in \texttt{R} for
computations.

Table
%TCIMACRO{\TeXButton{reftabTobit}{\ref{tabTobit}} }%
%BeginExpansion
\ref{tabTobit}
%EndExpansion
displays inferences about $\tau$, the effect of a twin on hours worked or,
more precisely, on the work fraction. For $\Gamma=1$, Table
%TCIMACRO{\TeXButton{reftabTobit}{\ref{tabTobit}} }%
%BeginExpansion
\ref{tabTobit}
%EndExpansion
displays randomization inferences assuming the differential comparison of
twins versus different-single-sex-child is free of bias from unmeasured
covariates. For $\Gamma>1$, sensitivity to unmeasured bias is displayed.
The point estimate of $\tau$ in the absence of bias is 0.0793 or about
8\% reduction in work hours ($0.08\times40=3.2$ hours per week) for a
mother with twins. More precisely, this is an 8\% reduction in work
fraction or a reduction of 3.2 hours per week for any mother who would work
at least 3.2 hours if she did not have twins. The results are insensitive
to small biases, say, $\Gamma\leq1.2$, but are sensitive to moderate
bias, $%
\Gamma=1.25$; however, we do not expect much bias in the differential
comparison. As noted in Section~\ref{ssDifRiskMatch} and \citet
{RosSil09}, in a matched pair, treatment-versus-control comparison, a
bias $%
\Gamma=1.25$ is produced by an unobserved covariate that doubles the odds
of treatment and doubles the odds of a positive treatment-minus-control pair
difference in outcomes.

%
%t3 #&#
\begin{table}
\caption{Inference about the Tobit effect $\tau$. For each $\Gamma$,
the sensitivity analysis gives the
maximum possible $P$-value testing the null hypothesis of no treatment
effect, $H_{0}\dvtx\tau=0$, the minimum
one-sided 95\% confidence interval and the minimum possible point
estimate. Inferences use the mean, but
$M$-estimates with Huber weights produced similar results}
\label{tabTobit}\label{balance2}
\begin{tabular*}{\textwidth}{@{\extracolsep{\fill}}lccc@{}}
\hline
$\bolds{\Gamma}$ & $\bolds{P}$\textbf{-value} &\textbf{95\% CI} & \textbf{Estimate} \\
\hline
1.0 & $1.6 \times10^{-13}$ & $\tau\ge0.0616$ & 0.0793 \\
1.1 & $2.0 \times10^{-6}$ & $\tau\ge0.0324$ & 0.0502 \\
1.2 & 0.0148 & $\tau\ge0.0058$ & 0.0237\\
1.25 & 0.1512 & & \\
\hline
\end{tabular*}
\end{table}

\begin{figure}

\includegraphics{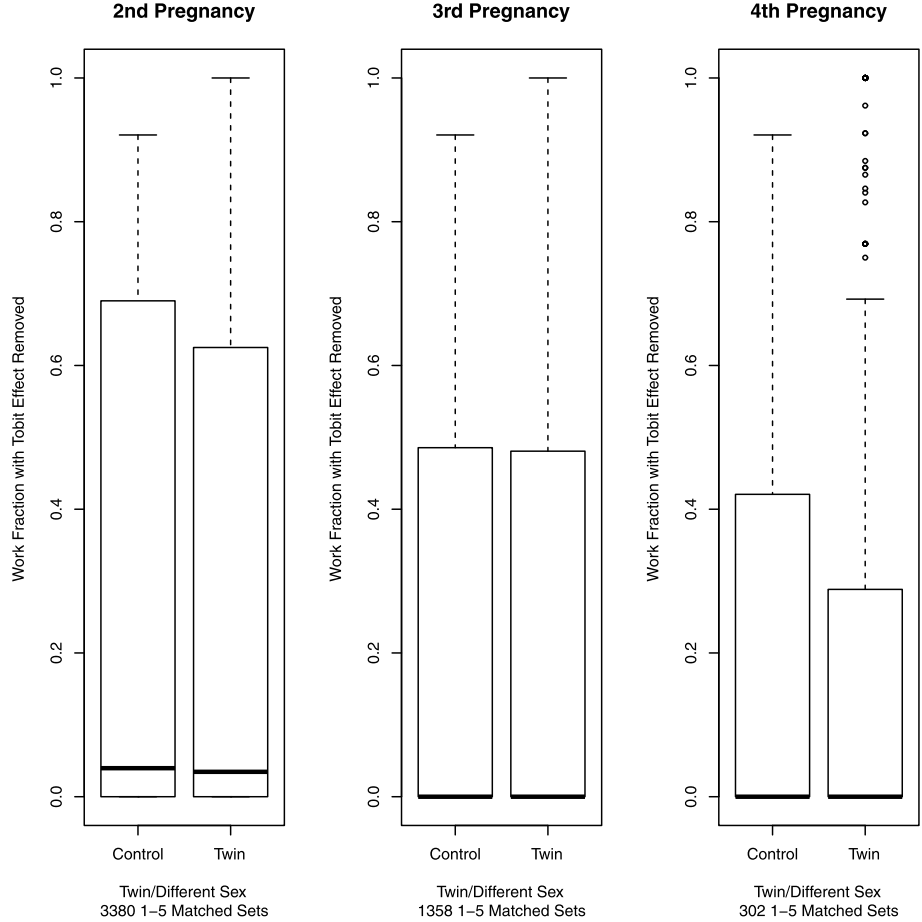}

  \caption{Residuals from the Tobit effect model. The boxplots display
$\max\{0, R_{ij}-(1 - Z_{ij})\tau_{0}\}$ for $\tau_{0} =
0.0793$, the point estimate of $\tau$ at $\Gamma=1$. In an infinitely large
sample, if the Tobit model were true with this $\tau$ and $\Gamma$, then
the pair of boxplots at each pregnancy would be identical.}\label{fig5}
\end{figure}

Figure~\ref{fig5} looks at residuals. With $\tau_{0}=0.0793$, Figure~\ref{fig5} plots
$\max
\{ 0, R_{ij}- ( 1-Z_{ij} ) \tau_{0} \} $. In an
infinite sample without bias, this plot would have identical pairs of
boxplots if the Tobit effect were correct. Though not identical in pairs,
the boxplots are similar, except perhaps at the 4th pregnancy where the
sample size is not large. Arguably, the data do not sharply contradict a
Tobit effect.

The second model related the effect on workforce participation to the effect
on the number of children, that is, the two outcomes in Figure~\ref{fig3}. Write
$%
D_{ij}$ for the number of children, with $D_{ij}=d_{Tij}$ if $Z_{ij}=1$
and $%
D_{ij}=d_{Cij}$ if $Z_{ij}=0$. The second model says the effect of
twin-versus-different-sex-single child on the workforce outcome is
proportional to the effect on the number of children,
$r_{Tij}-r_{Cij}=\beta
( d_{Tij}-d_{Cij} ) $. Under this model, $R_{ij}-\beta
D_{ij}=r_{Tij}-\beta d_{Tij}=r_{Cij}-\beta d_{Cij}$ does not change
with $%
Z_{ij}$, so (i) the null hypothesis $H_{0}\dvtx\beta=\beta_{0}$ is
tested by
testing the hypothesis of no effect of the treatment $Z_{ij}$ on $%
R_{ij}-\beta_{0}D_{ij}$, (ii) a confidence interval for $\beta$ is
obtained in the usual way by inverting the test, and (iii) a sensitivity
analysis for biased $Z_{ij}$ is conducted in the usual way; see
\citet{Ros96} and Imbens and Rosenbaum (\citeyear{ImbRos05}). This model embodies
the exclusion
restriction in saying that if the twin did not alter the total number of
children for mother $ij$, so $d_{Tij}=d_{Cij}$, then it did not alter her
workforce participation, $r_{Tij}=r_{Cij}$. For instance, if mother $ij$
had a twin on her second birth, $Z_{ij}=1$, she might have three
children, $%
d_{Tij}=3$, where perhaps she would have had two children if she had
had a
different-sex-single child at the second birth, $d_{Cij}=2$, so for this
mother the twin causes a 1 child increase in her number of children, $%
d_{Tij}-d_{Cij}=1$, and hence a change in workforce participation of $%
r_{Tij}-r_{Cij}=\beta ( d_{Tij}-d_{Cij} ) =\beta$. Some other
mother, $i^{\prime}j^{\prime}$, might have had three children
regardless, $%
d_{Tij}=d_{Cij}=3$, in which case the twin caused no increase in her number
of children, $d_{Tij}-d_{Cij}=0$ so $r_{Tij}-r_{Cij}=0$. \citet
{Baietal10} show that randomization inferences (i.e., inferences with
$\gamma
=\phi_{s}-\phi_{s^{\prime}}=0$) for $\beta$ under the model $%
r_{Tij}-r_{Cij}=\beta ( d_{Tij}-d_{Cij} ) $ are identical to
randomization inferences for the effect ratio, $ (
\sum_{i=1}^{I}\sum_{j=1}^{J}r_{Tij}-r_{Cij} ) / (
\sum_{i=1}^{I}\sum_{j=1}^{J}d_{Tij}-d_{Cij} ) $, which is the effect on
workforce participation per added child, and this is true whether or
not the
exclusion restriction holds. For instance, $\beta=-0.1$ would be a 0.1
reduction in the average work fraction per additional child, whether or
not $%
r_{Tij}-r_{Cij}=\beta ( d_{Tij}-d_{Cij} ) $ for each individual $%
ij $. Without the model $r_{Tij}-r_{Cij}=\beta (
d_{Tij}-d_{Cij} ) $, but with the exclusion restriction, the effect
ratio can be interpreted as the average effect on workforce participation
per child among mothers who had additional children because of the
twin; see
\citet{AngImbRub96}.

%t4 #&#
\begin{table}
\caption{Inference about the proportional effect, $\beta$. For each
$\Gamma$, the sensitivity analysis gives the
maximum possible $P$-value testing the null hypothesis of no treatment
effect, $H_{0}\dvtx\beta=0$, the minimum
one-sided 95\% confidence interval and the minimum possible point
estimate. Inferences use the mean, but
$M$-estimates with Huber weights produced similar results}
\label{tabeffratio}\label{balance2}
\begin{tabular*}{\textwidth}{@{\extracolsep{\fill}}lccc@{}}
\hline
$\bolds{\Gamma}$ & $\bolds{P}$\textbf{-value} & \textbf{95\% CI} & \textbf{Estimate} \\
\hline
1.0 & $1.6 \times10^{-13}$ & $\beta\le-0.0365$ & $-0.0470$ \\
1.1 & $2.0 \times10^{-6}$ & $\beta\le-0.0191$ & $-0.0296$ \\
1.2 & 0.0148 & $\beta\le-0.0034$ & $-0.0139$ \\
1.25 & 0.1512 & & \\
\hline
\end{tabular*}
\end{table}

Table
%TCIMACRO{\TeXButton{reftabeffratio}{\ref{tabeffratio}} }%
%BeginExpansion
\ref{tabeffratio}
%EndExpansion
draws inferences about the proportional effect, $\beta$. The test of no
treatment effect is the same as in Table
%TCIMACRO{\TeXButton{reftabTobit}{\ref{tabTobit}}}%
%BeginExpansion
\ref{tabTobit}%
%EndExpansion
, so the $P$-values in the two analyses are equally sensitive to unmeasured
biases. In the absence of unmeasured bias,\vadjust{\goodbreak} $\Gamma=1$, the point estimate
of $\beta$ suggests a 5\% reduction in the work fraction per additional
child.
We have been looking at the effects of twins versus the popular mix of
children of both sexes. The effects appear to be small.

%s5 #&#
\section{\texorpdfstring{Discussion.}{Discussion}}\label{sec5}

Isolation, as we have defined it, is used in the following situation. One
of several treatments may be inflicted upon individuals (or self-inflicted)
at certain moments in time. The timing $t$ of treatment may be severely
biased by both measured and unmeasured time-varying covariates, but there
may be two treatments, $s$ and $s^{\prime}$, such that conditionally given
some treatment at $t$, the occurrence of treatment $s$ in lieu of
treatment $%
s^{\prime}$ is close to random. Isolation focuses attention on that brief
moment and random aspect by controlling for measured time-dependent
covariates using risk-set matching and by removing a generic bias using a
differential comparison. Stated precisely, isolation refers to the radical
simplification of the conditional probability in (\ref{eqSenMod}) that
occurs when $\phi_{s}=\phi_{s^{\prime}}$; then, the unobserved time
dependent covariate $u_{ijt}$ that would bias most comparisons does not bias
a risk-set match of treatment $s$ in lieu of $s^{\prime}$. This radical
simplification, when it occurs, justifies one very specific analysis: the
comparison of matched sets with similar observed histories to time $t$ where
some individual received treatment $s$ and the rest received treatment $
s^{\prime}$. In the case study, the timing of births is biased by a
woman's plans and aspirations for education, career and family, but
conditionally given a birth at time $t$, the occurrence of twins rather than
a single birth is largely unaffected by her plans.

In a different study that employed similar reasoning, \citet
{NagSno13} examined the effects of incarceration on subsequent criminal
activity. The substantial difficulty is that judges decide in a thoughtful
manner whether to imprison an individual convicted for a crime. When two
people are convicted of the same crime, it is far from a random event when
one is sent to prison and the other is punished in a different way. Nagin
and Snodgrass looked at counties in Pennsylvania in which some judges were
much harsher than others, sending many more convicts to prison. Committing
a crime is not haphazard, nor is a judge's decision, but having your case
come to trial when judge A rather than judge B is next available is, in most
instances, a haphazard event. Nagin and Snodgrass contrasted the
subsequent criminal activity of individuals with similar pasts who were
tried before harsh judges and those tried before lenient judges in the same
county at about the same time, so each convict might have received either
judge. They found little or no evidence in support of the widespread
belief that harsher judges and harsher sentences reduce the frequency of
subsequent rearrest.

A similar strategy is sometimes used in studies of differential effects of
biologically different drugs used to treat the same disease. The
differential effect may be less confounded than the absolute effect of
either drug, particularly if the choice of drug is determined by something
haphazard. For example, \citet{Broetal06} compared the
gastrointestinal toxicity caused by COX-II inhibitors versus NSAIDs by
comparing the patients of physicians\vadjust{\goodbreak} who usually prescribe one versus those
who usually prescribe the other. See also \citet{Gibetal10} and
\citet{Ryaetal12}.

\printaddresses
\end{document}